\documentclass[prb,showpacs,twocolumn,amsmath]{revtex4}
\usepackage{color}
\usepackage{graphicx}
\usepackage{dcolumn}
\usepackage{bm}

\begin{document}

\preprint{FLEX}

\title{Doping dependence of spin fluctuations and electron correlations in iron pnictides}%

\author{Hiroaki Ikeda$^{1,2}$}%
\email{hiroaki@scphys.kyoto-u.ac.jp}
\author{Ryotaro Arita$^{2,3,4}$}%
\author{Jan Kune\v{s}$^{5}$}%
\affiliation{$^1$Department of Physics, Kyoto University, Kyoto, 606-8502, Japan\\
$^2$JST, TRIP, Sanbancho, Chiyoda, Tokyo 102-0075, Japan\\
$^3$Department of Applied Physics, University of Tokyo, Tokyo 113-8656, Japan\\
$^4$JST, CREST, Hongo, Tokyo 113-8656, Japan CREST\\
$^5$Institute of Physics,
Academy of Sciences of the Czech Republic, Cukrovarnick\'a 10,
162 53 Praha 6, Czech Republic
}%

\date{\today}%

\begin{abstract}
Doping dependence of the spin fluctuations and the electron correlations in the effective five-band Hubbard model for iron pnictides is investigated using the fluctuation-exchange approximation.
For a moderate hole doping, we find a dominant low-energy spin excitation at $\bm{Q}=(\pi,0)$, which becomes critical at low temperature.
The low-energy spin excitations in the heavily hole-doped region are characterized by weak $\bm{Q}$ dependence.
The electron doping leads to an appearance of a pseudogap in spin-excitation spectrum.
Correspondingly, the NMR-$1/T_1$ relaxation rate is strongly enhanced on the hole-doped side and suppressed on the electron-doped side of the phase diagram.
This behavior can be to large extent understood by systematic changes of the Fermi-surface topology.
\end{abstract}


\maketitle

\section{Introduction}
The recent discovery of iron-pnictide superconductors with high transition temperatures ($T_c$),~\cite{rf:Kamihara,rf:Takahashi} some over $50$K,~\cite{rf:Ren} has provoked an intense research.
The appearance of high-$T_c$ superconductivity in a close proximity to the anti-ferromagnetic (AF) phase, induced by a carrier doping and an applied pressure, reminds of the cuprates and the heavy-fermion superconductors.
The sign-reversing $s_\pm$-wave state with the pairing glue of AF fluctuations has been suggested as an explanation of the observed superconductivity.~\cite{rf:Mazin,rf:Kuroki,rf:Yanagi,rf:Ikeda,rf:Graser,rf:Nomura,rf:FWang,rf:Yao}
The NMR~\cite{rf:Nakai,rf:Grafe,rf:Ahilan,rf:Mukuda,rf:Matano,rf:Fukazawa,rf:Yashima,rf:Mukuda2,rf:Ning,rf:Zhang,rf:Fukazawa2} and the neutron-scattering experiments~\cite{rf:Christianson,rf:Wakimoto,rf:Inosov,rf:Matan,rf:Wang} revealed in detail the evolution of magnetic excitations with the carrier doping, in particular the presence of the strong AF fluctuations in the hole-doped systems and opening of a (pseudo)gap for the electron doping.
The uniform susceptibility has shown puzzling linear increase over a broad temperature range, which does not match with the two common limits, $T$-independent for weakly interacting electrons and $1/T$ for local moments.
The effect of electron correlations has been observed and quantified by means of the angle-resolved photoemission spectroscopy (ARPES),~\cite{rf:Lu} the optical spectroscopy,~\cite{rf:Qazilbash} and de Haas-van Alphen (dHvA) experiments.~\cite{rf:Shishido,rf:Terashima}
The accumulated knowledge suggests that understanding the changes of spin dynamics and electronic correlations with the carrier doping provides the key to the superconducting-paring mechanism.

Previously, one of us (H.I.) studied the doping and temperature dependence of the single-particle spectra and the NMR-$1/T_1$, applying the fluctuation-exchange (FLEX) approximation to an effective five-band Hubbard model, obtained for the band structure of LaFeAsO.~\cite{rf:Ikeda}
This early study succeeded in predicting the enhancement of the AF spin fluctuations on the hole-doped side, 
but also revealed a problem with double counting the interaction effects in the multiband systems. 
Straightforward addition of the FLEX self-enenrgy leads to redistribution of orbital occupancies 
from their LDA values accompanied by drastic changes in the Fermi surface and the spin fluctuations in
contrast to experimental observations.
In the preceding papers,~\cite{rf:Arita,rf:Ikeda2} we have investigated this point in detail and proposed a way around,
subtraction of a static part of the self-energy, together with physical argument supporting this {\it ad hoc} procedure.
With this modification, we now can apply the FLEX approximation over wide ranges of carrier doping and temperature.

Here we employ the modified FLEX scheme~\cite{rf:Ikeda2} to study the effect of varying electron concentration on spin-fluctuations in the five-band model of iron pnictides and compare it with the experimentally observed trends across the pnictide series.
In order to single out the effect of carrier doping we do not construct precise tight-binding models 
for each individual compound, but use the parameters obtained for LaFeAsO.~\cite{rf:fs}
The calculations capture the common features in the series of iron-pnictide superconductors derived from LaFeAsO 
and BaFe$_2$As$_2$.

\section{Model and computational details}
We start with a five-band Hubbard model in the unfolded two-dimensional Brillouin zone (BZ).
The kinetic term 
\begin{equation}
H^0=\sum_{\bm{k}\ell m \sigma}h^{\bm{k}}_{\ell m}c_{\bm{k}\ell\sigma}^\dagger c_{\bm{k}m\sigma}
\end{equation}
comes from the tight-binding model in LaFeAsO, and the hopping integrals appear in TABLE II of Ref.~\onlinecite{rf:Ikeda2}.
Only the on-site Coulomb interaction is considered with the common parametrization:
the intraorbital Coulomb $U$, the interorbital Coulomb $U'$, the Hund coupling $J$, and the pair-hopping $J'$.
In the FLEX approximation, the normal Green's functions $\mathcal{G}_{\ell m}(\bm{k},i\omega_n)$ for orbitals $\ell$ and $m$ are self-consistently determined from the following equations,
\begin{subequations}
\begin{align}
& \mathcal{G}_{\ell m}(k) =\mathcal{G}^0_{\ell m}(k)
+\sum_{\ell'm'}\mathcal{G}^0_{\ell\ell'}(k)\Sigma_{\ell'm'}(k)\mathcal{G}_{m'm}(k), \\
&\Sigma_{\ell m}(k) =\sum_q\sum_{\ell'm'} V_{\ell\ell',mm'}(q)\mathcal{G}_{\ell'm'}(k-q), \\
&V_{\ell\ell',mm'}(q) =\Big[\hat{U}_{\uparrow\downarrow}-2\hat{U}_{\uparrow\uparrow}
-\hat{U}_{\uparrow\downarrow}\hat{\chi}^0(q)\hat{U}_{\uparrow\downarrow} \nonumber \\
& \hspace{45pt}+\frac{3}{2}\hat{U}^s\hat{\chi}^s(q)\hat{U}^s+\frac{1}{2}\hat{U}^c\hat{\chi}^c(q)\hat{U}^c
\Big]_{\ell\ell',mm'}.
\end{align}
\end{subequations}
Here the bare vertices $\hat{U}^{s,c}=\hat{U}_{\uparrow\downarrow}\mp\hat{U}_{\uparrow\uparrow}$ with $(\hat{U}_{\uparrow\downarrow})_{\ell\ell,\ell\ell}=U$, $(\hat{U}_{\uparrow\downarrow})_{\ell\ell,mm}=U'$, $(\hat{U}_{\uparrow\downarrow})_{\ell m,\ell m}=J$, $(\hat{U}_{\uparrow\downarrow})_{\ell m,m\ell}=J'$, $(\hat{U}_{\uparrow\uparrow})_{\ell\ell,mm}=U'-J$, $(\hat{U}_{\uparrow\uparrow})_{\ell m,\ell m}=J-U'$ for $\ell\ne m$, and the susceptibilities in the spin sector and the charge sector are given by
\begin{subequations}
\begin{align}
& \hat{\chi}^s(q) =\hat{\chi}^0(q)+\hat{\chi}^0(q)\hat{U}^s\hat{\chi}^s(q), \\
& \hat{\chi}^c(q) =\hat{\chi}^0(q)-\hat{\chi}^0(q)\hat{U}^c\hat{\chi}^c(q),
\end{align}
\end{subequations}
and
\begin{equation}
\big[\hat{\chi}^0(q)\big]_{\ell\ell',mm'}=-\sum_k \mathcal{G}_{\ell m}(k+q)\mathcal{G}_{m'\ell'}(k).
\end{equation}
With the straightforward application of FLEX, we encounter the double counting problem mentioned in the
Introduction.
Therefore, as in Ref.~\onlinecite{rf:Ikeda2}, we subtract the $\omega=0$ part of the self-energy in order 
to eliminate the unwanted redistribution of the charge between orbitals.  
In Ref.~\onlinecite{rf:Ikeda2} we argued that this mimics the effect of the Hartree part of 
the electron-electron interaction, 
which does not appear in our low-energy effective model. The correction term $\Sigma_{\ell m}^R(\bm{k},\omega=0)$ 
was calculated at $T=23K$ (Ref.~\onlinecite{rf:limit}) and used unchanged also at higher temperatures so that 
the low-temperature Fermi surface matches the LDA one, but at higher temperature small 
modifications of the Fermi surface are allowed.~\cite{rf:details}
Once the correction term is fixed, this procedure equals the standard FLEX approximation 
for a Hamiltonian with the kinetic part, 
$\mathcal{H}^{\bm{k}}_{\ell m}=h^{\bm{k}}_{\ell m}-\Sigma_{\ell m}^R(\bm{k},0)$
instead of $h^{\bm{k}}_{\ell m}$.

Through this paper, we take $64\times 64$ meshes in the unfolded BZ and $1024$ Matsubara frequencies, and $U=1.20$eV and $J=0.25$eV as the interactions with $U=U'+2J$ and $J'=J$.
The retarded quantities $\mathcal{G}_{\ell m}^R(\bm{k},\omega)$, $\Sigma_{\ell m}^R(\bm{k},\omega)$ and $\chi_s^R(\bm{q},\omega)$ are obtained by the numerical analytic continuation with use of the Pad\'e approximation.

\section{Results and discussion}
\subsection{Single-particle spectra and Fermi surface}
\begin{figure}
\centering
\vspace{5pt}
\includegraphics[width=85mm]{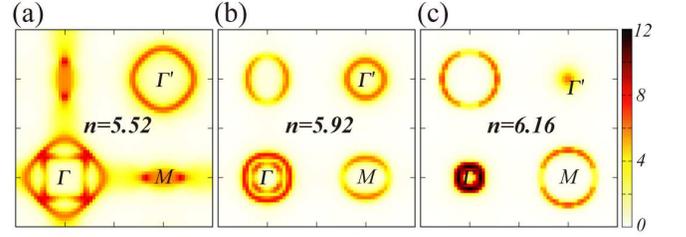}
\caption{(Color online) Contour plots of the Green's function, $-\sum_{\ell}$Im$\mathcal{G}_{\ell\ell}(\bm{k},i\pi T)/\pi$, at $T=4$meV$\simeq 46$K for $n=5.52$ (a), $5.92$ (b), and $6.16$ (c) in the unfolded BZ. Deep red corresponds to the Fermi surface. The Fermi surface around $M$ point is the electron sheet at $n=5.92$ and $6.16$, but the hole sheet at $n=5.52$. The weight around $\varGamma'$ point becomes small at lower temperatures, since the band edge locates below the Fermi level.}
\label{fig:fs}
\end{figure}
The Fermi Surfaces (FS) formed by the renormalized bands for fillings~\cite{rf:filling} corresponding to heavy, $n=5.52$,  and moderate, $n=5.92$, hole-doping and heavy electron-doping, $n=6.16$
are shown in Fig.~\ref{fig:fs}. 
The broadening comes from the imaginary part of the self-energy, arising mainly from the magnetic scattering.
The $n=5.92$ FS consists of hole sheets around the $\varGamma$ and the $\varGamma'$ points and 
an electron sheet centered at the $M$ point nested with the well known vectors $\bm{Q}=(\pi,0)$ or $(0,\pi)$.
Electron doping leads, as expected, to shrinking of the $\varGamma$ ($\varGamma'$) hole sheets and expansion of the $M$ point electron surface, as shown in Fig.~\ref{fig:fs}c.
At  $n=6.16$ the $\varGamma'$ surface is reduced to almost a point, which leads to a $T$ 
dependence discussed below.
The $n=5.52$ FS, similar to that of KFe$_2$As$_2$, has large hole sheets around the $\varGamma$ ($\varGamma'$) points and a small hole pocket around the $M$ point.

\begin{figure}
\centering
\includegraphics[width=85mm]{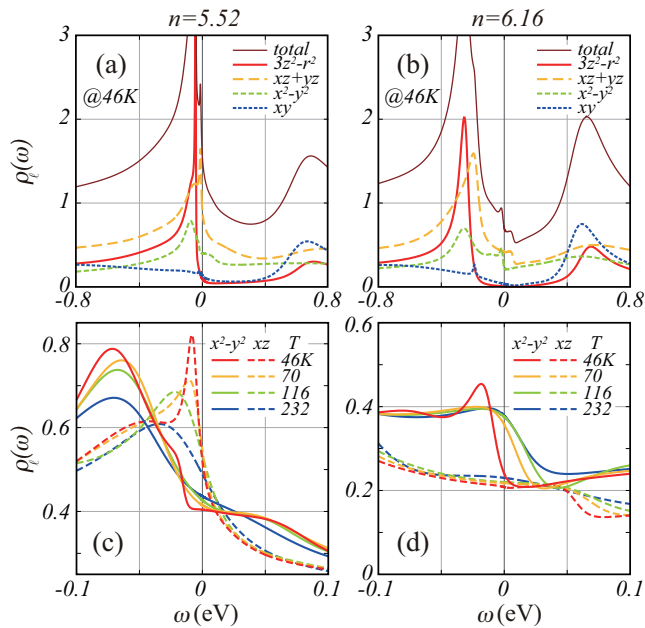}
\caption{(Color online) The partial density of states $\rho_\ell(\omega)$ (states/eV) at $n=5.52$ (a) and $n=6.16$ (b) at $46$K. Temperature dependence for $d_{xz/yz}$ and $d_{x^2-y^2}$ orbitals at $n=5.52$ (c) and $n=6.16$ (d). Shown is remarkable suppression of $\rho_\ell(0)$ for $d_{x^2-y^2}$ orbital at $n=6.16$.}
\label{fig:dos}
\end{figure}
Figs.~\ref{fig:dos}a and \ref{fig:dos}b show the orbital-resolved spectral densities 
$\rho_\ell(\omega)=-\sum_{\bm k}$Im$\mathcal{G}^R_{\ell\ell}(\bm{k},\omega)/\pi$ for $n=5.52$ and $6.16$.
Like for the non-interacting bands, the Fermi level $E_{\rm F}=0$ falls to a vicinity of a spectral peak, more so for the hole doped systems.
The states on the Fermi surface arise predominantly from $d_{xz/yz}$ and $d_{x^2-y^2}$ orbitals.
Their relative weight varies strongly with doping and while on the hole-doped side the $d_{xz/yz}$ contribution dominates over $d_{x^2-y^2}$, the reverse holds for the electron-doping.

Figs.~\ref{fig:dos}c and \ref{fig:dos}d show the $T$-dependent changes of the spectral functions in the vicinity of the Fermi level.
A remarkable suppression of the $d_{x^2-y^2}$ density at $E_{\rm F}$ upon cooling can be seen for $n=6.16$, when  the Fermi level is located close to a step-like van Hove singularity at the top of the $\varGamma'$ band.
The decreasing temperature leads not only to reduced quasi-particle damping, but also to a downward band shift  and the corresponding shrinking of the $\varGamma'$ pocket.
Such a variation of the spectral density has been observed in the photoemission spectra,~\cite{rf:Sato} and numerical study of Ref.~\onlinecite{rf:Ikeda}.
The shrinking of FS has been also observed in LaFePO (Ref.~\onlinecite{rf:Lu}) and BaFe$_2$(As$_{1-x}$P$_x$)$_2$ (Ref.~\onlinecite{rf:Shishido}). 
General arguments for correlation-induced band shifts due to coupling of an asymmetric electronic band to a bosonic mode, e.g. spin fluctuation, were given by Ortenzi {\it et al.}~\cite{rf:Ortenzi}
Applying their reasoning together with the increase of spin-fluctuation density at low-$T$ 
(Fig.~\ref{fig:neutron}), indeed, leads to the observed shift of the $\varGamma'$ band.

\begin{figure}
\centering
\includegraphics[width=85mm]{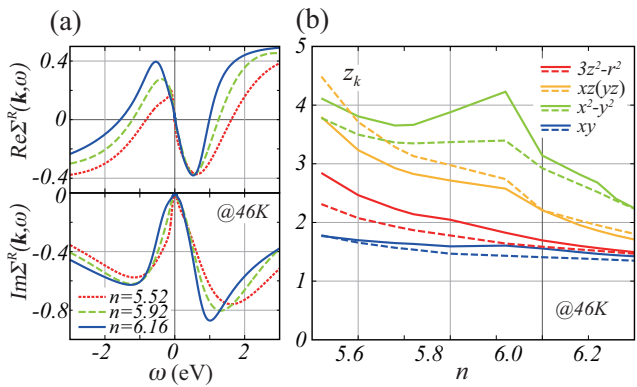}
\caption{(Color online) (a) The self-energy for $d_{x^2-y^2}$ orbital at the $\varGamma'$ point at $46$K in units of eV. (b) Mass enhancement factor $z^\ell_{\bm k}$ for each orbital at $\bm{k}=(26\pi/32,0)$ around the $M$ point (solid) and $\bm{k}=(6\pi/32,0)$ around the $\varGamma$ point (dotted). }
\label{fig:zk}
\end{figure}

The main effect of electronic correlations in itinerant systems is an enhancement of the quasiparticle mass.
In Fig.~\ref{fig:zk}a, we use typical self-energies to demonstrate that the scattering processes contained in FLEX lead to quasiparticle renormalization over a rather broad energy range. In fact, we find a fair agreement between the FLEX self-energy and the self-energy obtained with LDA+DMFT for the five band model.~\cite{rf:Anisimov}
This applies also to the mass enhancement factors, except for the region around $n\sim 5.92$ where nesting, and thus strongly ${\bm k}$-dependent effects, dominate the physics.
In Fig.~\ref{fig:zk}b, we show the doping dependence of the mass enhancement factors $z^\ell_{\bm k}$ for each orbital,
\begin{equation}
z^\ell_{\bm k}=1-\frac{\partial\varSigma^R_{\ell\ell}(\bm{k},\omega)}{\partial\omega}\biggr|_{\omega\to 0}
\simeq 1-\frac{{\rm Im}\varSigma_{\ell\ell}(\bm{k},i\pi T)}{\pi T},
\end{equation}
in the vicinity of the $\varGamma$ and $M$ sheets of the Fermi surface.  
We observe an overall trend of increasing mass enhancement from $2-3$ on the electron-doped side to over 4 on the hole-doped side, which can be attributed to a growing spectral  density at $E_F$ (see Fig.~\ref{fig:dos}a).
In agreement with this trend, a remarkably large mass enhancement factor is observed in dHvA on KFe$_2$As$_2$.~\cite{rf:Terashima}
As shown below, this can be because low-energy spin fluctuations become featureless, $\bm{Q}$-independent, similar to the heavy fermion systems.
The cusps in $z^\ell_{\bm k}$ for $d_{x^2-y^2}$ orbital at $n\sim 5.92$ result from strong stripe-type AF spin fluctuations due to the FS nesting.

\subsection{Spin-excitation spectra}
\begin{figure*}
\centering
\includegraphics[width=\textwidth]{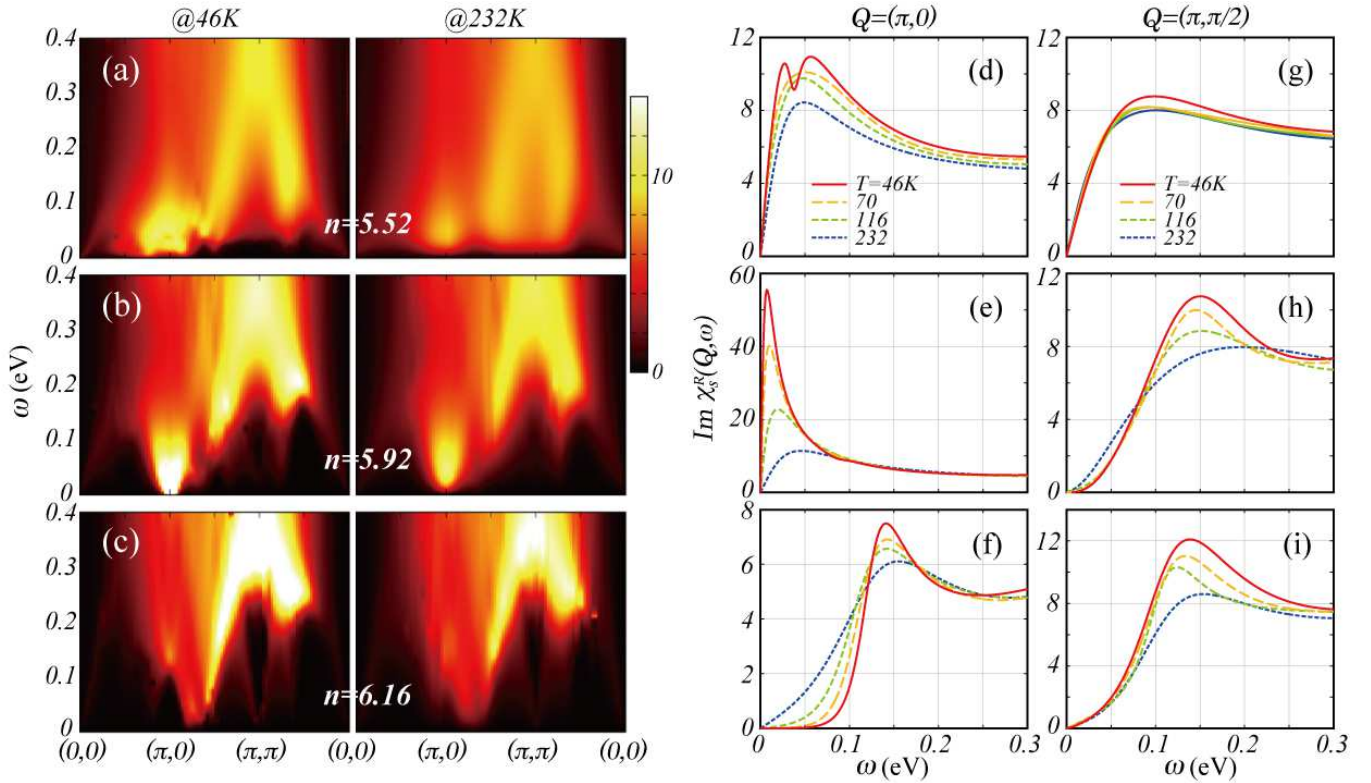}
\caption{(Color online) Imaginary part of $\chi_s^R(\bm{Q},\omega)$ ($\mu_{\rm B}^2/{\rm eV}$) along the high-symmetry line at $n=5.52$ (a), $n=5.92$ (b), and $n=6.16$ (c). Temperature dependence at $\bm{Q}=(\pi,0)$ and $(\pi,\pi/2)$ at $n=5.52$ (d, g), $n=5.92$ (e, h), and $n=6.16$ (f, i). The system at $n=5.92$ locates at around the stripe-type AF critical point. At $n=5.52$, shown is $\bm{Q}$-independent low-energy excitation, while the gap-like behavior at $n=6.16$.}
\label{fig:neutron}
\end{figure*}
Next, we discuss the structure of the spin excitation, Im$\chi_s^R(\bm{Q},\omega)$, at $n=5.52$, $5.92$, and $6.16$, displayed in Fig.~\ref{fig:neutron}.
In the left panel we show contour plots of  Im$\chi_s^R(\bm{Q},\omega)$ along the high symmetry directions.
The right panel shows the variation of  Im$\chi_s^R(\bm{Q},\omega)$ at fixed $\bm{Q}$ with temperature. 

For a moderate hole doping, $n=5.92$, the strong low-energy fluctuations (stronger than in the undoped case), arising from FS nesting, are located around $\bm{Q}=(\pi,0)$ (see Fig.~\ref{fig:neutron}b).
They exhibit strong enhancement at low $T$ indicating an incipient AF instability (Fig.~\ref{fig:neutron}e).
Further hole doping destroys the nesting between $\varGamma$ ($\varGamma'$) and $M$ sheets of FS, and increases the density of states at the Fermi level as shown in Fig.~\ref{fig:dos}a.
This leads to rather featureless, $\bm{Q}$-independent, and only weakly $T$-dependent structure of the low-energy spin fluctuations as shown in Figs.~\ref{fig:neutron}a, \ref{fig:neutron}d, and \ref{fig:neutron}g. 
A broad hump around $\bm{Q}=(\pi,0)$, which develops at low $T$, 
is due to scattering between the $\varGamma$ hole sheets and the $M$ pocket, composed of $d_{xz/yz}$ orbitals, in Fig.~\ref{fig:fs}a.

In contrast, the electron doping results in suppression of the low-energy spin fluctuations.
For $n=6.16$ the $(\pi,0)$ spin fluctuations are remarkably suppressed, and a gap opens in the spin excitation spectrum at low $T$ as shown in Figs.~\ref{fig:neutron}c and \ref{fig:neutron}f.
Recently, such a gap-like behavior with heavy electron doping has been observed by the inelastic neutron-scattering experiment.~\cite{rf:Matan}
This behavior is linked to the corresponding changes of the Fermi-surface topology in Fig.~\ref{fig:fs}c.
As the electron doping leads to the shrinking of the $\varGamma$ ($\varGamma'$) hole sheets and expansion of the $M$ electron sheet, the particle-hole excitation at $\bm{Q}=(\pi,0)$ requires a finite energy, and the corresponding spectral weight moves to higher energies. 
The leading scattering channel becomes dominated by scattering between different electron ($M$) sheets, and the dominant low-energy spin fluctuation moves to $\bm{Q}=(\pi,\pi/2)$, as shown in Fig.~\ref{fig:neutron}i.
Previous calculations showed that the leading superconducting instability changes from $s_\pm$-wave to $d_{x^2-y^2}$-wave around this doping.~\cite{rf:Ikeda2}
This points to the correlation between the structure of the spin fluctuations and the superconducting pairing symmetry.

\subsection{NMR $1/T_1$ and uniform susceptibility}
\begin{figure}
\centering
\includegraphics[width=75mm]{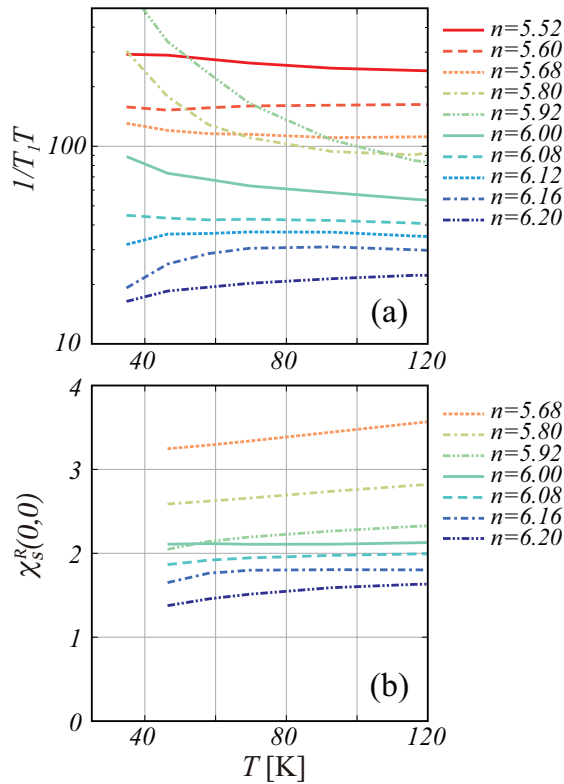}
\caption{(Color online) (a) NMR $1/T_1T$ as a function of temperature. The vertical line is a logarithmic scale. Shown are remarkable enhancement on the hole-doped side and suppression on the electron-doped side. (b) The uniform susceptibility $\chi_s^R(0,0)$ ($\mu_{\rm B}^2/{\rm eV}$), which overall decreases upon cooling.~\cite{rf:accuracy}}
\label{fig:chs}
\end{figure}

Next, we discuss the local spin response measured in terms of the NMR-$1/T_1$ relaxation rate.
Fig.~\ref{fig:chs}a shows the temperature dependence of 
$1/T_1T=\sum_{\bm q}$Im$\chi_s^R(\bm{q},\omega)/\omega|_{\omega \to 0}$ for various dopings with the hyperfine-coupling constant set to unity.
The overall trend of $1/T_1T$ growing with the hole doping at high temperatures follows from increasing $\rho_\ell(0)$.
At lower temperatures, $1/T_1T$ is suppressed on the electron-doped side, reflecting the reduction of $\rho_\ell(0)$, as shown in Fig.~\ref{fig:dos}d, and the corresponding opening of a spin gap (Fig.~\ref{fig:neutron}c and \ref{fig:neutron}f). 
Such suppression of $1/T_1T$ with electron doping, which is consistent with the previous works,~\cite{rf:Ikeda,rf:Laad} has been observed in LaFeAs(O$_{1-x}$F$_x$) (Ref.~\onlinecite{rf:Nakai,rf:Grafe,rf:Mukuda,rf:Ahilan}), LaFeAsO$_{1-y}$ (Ref.~\onlinecite{rf:Mukuda2}) and Ba(Fe$_{1-x}$Co$_x$)$_2$As$_2$ (Ref.~\onlinecite{rf:Ning}).
A remarkable growth of $1/T_1T$ in $n=5.92$ and $5.80$ cases comes from the enhanced AF spin fluctuation shown in Figs.~\ref{fig:neutron}b and \ref{fig:neutron}e.
Although such enhancement is suppressed with further the hole doping, a slight increase can be observed also for $n=5.52$, originating from the broad hump structure around $Q=(\pi,0)$ in Figs.~\ref{fig:neutron}a and \ref{fig:neutron}d.
The trends in $1/T_1T$ are consistent with a remarkable low-$T$ increase in Ba$_{0.72}$K$_{0.28}$Fe$_2$As$_2$ (Ref.~\onlinecite{rf:Matano}) and Ba$_{0.6}$K$_{0.4}$Fe$_2$As$_2$ (Ref.~\onlinecite{rf:Mukuda2,rf:Yashima,rf:Fukazawa}), and a relatively small increase in KFe$_2$As$_2$ (Ref.~\onlinecite{rf:Zhang,rf:Fukazawa2}) at low temperatures.

Finally, we present  $T$-dependence of the uniform susceptibility $\chi_s^R(0,0)$ in Fig.~\ref{fig:chs}b.
Roughly speaking, it exhibits an increase with temperature independent of the carrier doping except for $n=6.00$ case.
On the hole-doped side, the tendency is accompanied by the development of the AF fluctuation, in qualitative agreement with a scenario studied in  Ref.~\onlinecite{rf:Chubukov}.
On the electron-doped side, it is consistent with the suppression of the NMR-$1/T_1T$ and $\rho_\ell(0)$, which is not accompanied by a remarkable development of the magnetic fluctuation.
Experimental observations shown clear a temperature dependence on the electron-doped side in LaFeAsO$_{1-x}$F$_x$ and 
Ba(Fe$_{1-x}$Co$_x$)$_2$As$_2$,~\cite{rf:Ahilan,rf:Mukuda2,rf:Ning,rf:Klingeler,rf:XFWang} and a 
constant behavior on the hole-doped side in Ba$_{1-x}$K$_x$Fe$_2$As$_2$.~\cite{rf:Matano,rf:Yashima}
Thus, we need further investigations for complete understanding of the uniform susceptibility.

\section{Conclusions}
In conclusion, we have investigated the normal-state properties in the effective five-band Hubbard model for the iron-pnictides using the FLEX approximation.
We have obtained a variety of trends in the spin dynamics and the electron correlations with the carrier doping, which qualitatively agree with the overall features observed in the (Ba,K)Fe$_2$As$_2$, Ba(Fe,Co)$_2$As$_2$ and LaFeAsO systems, in particular, the gap-like feature in $(\pi,0)$ spin-excitation spectrum in the heavily electron-doped case, corresponding to the Ba(Fe$_{1-x}$Co$_x$)$_2$As$_2$ with large $x$, and the weak $\bm Q$ dependence of the spin fluctuation and the large mass enhancement in the heavily hole-doped case, corresponding to the end material KFe$_2$As$_2$.
We find that the changes in the Fermi-surface topology are the main driving force behind the observed trends.

\section*{Acknowldgements}
We are grateful to Y. Matsuda, T. Shibauchi, K. Ishida, Y. Nakai, H. Fukazawa, S. Kasahara, K. Hashimoto, H. Shishido, T. Takimoto, and K. Yamada for fruitful discussion.
This work is supported by a Grant-in-Aid for Scientific Research on Priority Areas (No. 20029014) and the Global COE Program ``The Next Generation of Physics, Spun from Universality and Emergence'' from the Ministry of Education, Culture, Sports, Science and Technology, Japan.


\begin{thebibliography}{99}
\bibitem{rf:fs}
Still, the Fermi-surface geometry for $n\simeq 5.5$ in Fig.~\ref{fig:fs}a is very similar to that of KFe$_2$As$_2$, which corresponds to $0.5$ hole doping per Fe.
\bibitem{rf:limit}
The lowest temperature $T=23$K comes from the limitation of $1024$ Matsubara frequencies.
More Matsubara frequencies are required for calculations at lower temperature.
\bibitem{rf:details} 
First, we carry out the FLEX with modified self-energy at $T=23$K, following Ref.~\onlinecite{rf:Ikeda2}, and obtain the proper self-energy shift in each doping case.
When we subtract the proper shift from the original tight-binding Hamiltonian $h^{\bm{k}}_{\ell m}$,
the remainder, $\mathcal{H}^{\bm{k}}_{\ell m}=h^{\bm{k}}_{\ell m}-\Sigma_{\ell m}^R(\bm{k},0)$, can be reconsidered the proper unperturbed Hamiltonian in each doping case.
We apply the standard FLEX calculation without subtraction of the self-energy shift to this proper unperturbed Hamiltonian at each temperature.
More precisely, $\Sigma_{\ell m}^R(\bm{k},0)=\bm{i}_{\bm{k}\ell}\tilde{\Sigma}^R_{\ell m}(\bm{k},0)\bm{i}_{\bm{k}m}^*$ should be $\bm{i}_{\bm{k}\ell}{\rm Re}\tilde{\Sigma}^R_{\ell m}(\bm{k},0)\bm{i}_{\bm{k}m}^*$, where $\bm{i}_{\bm{k}\ell}$ is a diagonal matrix to eliminate a trivial phase factor which appears in $h_{\ell m}^{\bm{k}}$, defined in Ref.~\onlinecite{rf:Ikeda2}.
\bibitem{rf:accuracy} 
We do not show data at $T=35$K in $\chi_s^R(0,0)$ due to doubtful accuracy, while an error in calculation of $1/T_1T$ is averaged due to the summation over $\bm{k}$.
\end{thebibliography}
\end{document}